Magnetic order and disorder in nanomagnets probed by superconducting vortices

V. Rollano[1], J. del Valle[2,3], A. Gomez[4], M. Velez[5,6], L. M. Alvarez-Prado[5,6], C. Quiros[5,6], J. I. Martin[5,6], M. R. Osorio[1], D. Granados[1], E. M. Gonzalez[1,2], and J. L. Vicent[1,2]

[1] IMDEA-Nanociencia, Cantoblanco, 28049 Madrid, Spain.

[2] Departamento Física de Materiales, Facultad CC. Físicas, Universidad Complutense, 28040 Madrid, Spain.

[3] Department of Physics, Center of Advance Nanoscience, University of California, San Diego, 9500 Gilman Drive, La Jolla, California 92093, USA.

[4] Centro de Astrobiología (CSIC-INTA), Torrejón de Ardoz, 28850 Madrid, Spain.

[5] Departamento de Física, Universidad de Oviedo, 33007 Oviedo, Spain

[6] CINN (Universidad de Oviedo-CSIC), 33940 El Entrego, Spain.

Correspondence to jlvicent@ucm.es

We have studied two nanomagnet systems with strong (Co/Pd multilayers) and weak (NdCo alloy films) stray magnetic fields by probing the out-of-plane magnetic states with superconducting vortices. The hybrid samples are made of array of nanomagnets embedded in superconducting Nb thin films. The vortex motion detects relevant magnetic state features, since superconducting vortices are able to discriminate between different magnetic stray field strengths and directions. The usual matching effect between the superconducting vortex lattice and the periodic pinning array can be quenched by means of disorder magnetic potentials with strong stray fields at random. Ordered stray fields retrieve the matching effect and yield asymmetry and shift in the vortex dissipation signal. Furthermore vortices can discriminate the sizes of the nanomagnet magnetic domains, detecting magnetic domain sizes as small as 70 nm. In addition, we observe that the vortex cores play the crucial role instead of the supercurrents around the vortex.

Superconductivity and magnetism are two long range order phenomena which do not seem to be compatible at first sight. In spite of this perspective, the coexistence of both cooperative effects in the same material was foreseen and studied long time ago [1]. A pioneering experimental work was reported by Matthias et al. [2]. Taking into account these results, Anderson and Shul [3] found the conditions for spin alignment in the superconducting state in the case of rare earth magnetic moments diluted in a superconductor. Years later, long range order magnetism and superconductivity were found in ternary rare earth compounds [4]. Magnetism with reentrant superconductivity behavior [5], as well as ferromagnetism and superconductivity coexistence [6] without reentrant behavior were reported in these ternary rare earth crystals. After the finding of these pioneer compounds many materials have been found where both cooperative phenomena are present. Very different compounds can be quoted, some examples are $RuSr_2GdCu_2O_8$ and $UGe_2$ [7, 8]. Nowadays, the most interesting finding is the iron-based compound family [9], as well as the coexistence and emergence of both phenomena in very peculiar sites, as for example interfaces [10]. In this scenario of interplay and coexistence between magnetism and superconductivity, another remarkable advance was foreseen by Matthias and Suhl [11]. These authors suggested that the superconducting regions extend only through the thicknesses of the ferromagnetic domain walls. This interesting idea was worked out and expanded by Tachiki et al. [12], Kulic [13] and Buzdin et al. [14, 15]. These authors found that the crucial parameter is the superconducting coherence length; they concluded that the comparison between this length and the domain wall or the domain size precludes or enhances the coexistence of superconductivity and magnetism.

Eventually, over the years the interplay between magnetism and superconductivity has evolved from competition to collaboration. The role played by magnetism in promoting superconductivity has called the attention of many researchers. In this way, the new scenario opened by magnetic/superconducting nanostructured hybrids has allowed tailoring superconducting properties almost at will by using the appropriate design of the nanomagnets. A simple example is to use applied magnetic fields to compensate, at the nanoscale, the stray fields coming from the nanomagnets. In this way, external magnetic fields promote superconductivity instead of preventing it [16-18] giving rise to the so-called magnetic-field-induced superconductivity. The combination of nanomagnets in contact with superconductors has generated a complete and mature field where researchers have been able to obtain remarkable achievements; a couple of examples could be control of vortex motion [19-22], and vortex-antivortex interplay [23-25]. In this work, we have fabricated arrays of nanomagnets embedded in superconducting films and we have taken the opposite approach, we have explored whether or not we can obtain information of the magnetic states of nanomagnets using the motion of superconducting vortices as a probe. Belkin et al. [26] have reported the influence of magnetic domain structure on vortex dynamics. In the present work, we are going one step further and we show that vortex lattice motion detects subtle differences linked to the strengths and configurations of the stray magnetic fields generated by nanomagnets. The increased pinning is due to weakened superconductivity. This weakened superconductivity is due to superconducting currents induced by the local magnetic field gradients at the domain walls and edges of nanomagnets [26, 27].

In nanomagnets, the domain sizes are usually governed by magnetostatic effects related to geometry and size of the magnetic nanoelement [28]. This gives rise to simple magnetic configurations, such as single domain, vortex state, etc, in which the domain size is comparable to the patterned element size. Stray fields created by each magnetic element and the order in the periodic array can be used to tune periodic pinning effects in the superconducting/magnetic hybrid [29]. An interesting but unexplored option would be to reduce magnetic domains below typical patterned element size (100 nm to 1000 nm). This can be done with materials which exhibit perpendicular magnetic anisotropy (PMA). Domain size can be varied from a few nm to several μm with a large variety of geometrical configurations (parallel stripes, labyrinths, bubbles, single domain …) depending on the interplay between material parameters (saturation magnetization $M_S$, anisotropy $K_N$ and disorder), element geometry (thickness and lateral dimensions) and magnetic history [22-30]. In a film, domain configuration is governed by Q, the ratio of perpendicular magnetic anisotropy $K_N$ to magnetostatic energy $Q= 2K_N/\mu_0 M_S^2$ and by the effect of disorder on nucleation and propagation processes of reversed domains [31]. High Q values (of the order 2 to 10) can be found, for example, in Co/Pd multilayers [32-35] with high out-of-plane remanence ($M_R \approx M_S$) and large domain sizes during magnetization reversal [36]. A smaller domain size regime can be accessed in lower Q materials such as NdCo amorphous alloys (Q in the 0.1-1.5 range, depending on temperature and composition) [37, 38]. These films display reduced remanent magnetization ($M_R$ below $0.1 M_S$) characteristic of stripe domain patterns, with very small domain sizes [39]. In summary, Co/Pd multilayers and (NdCo) films exhibit very different magnetic scenarios, since the former shows well defined and strong perpendicular magnetic anisotropy (high values of Q) and large domain sizes, and the latter shows a weak perpendicular magnetic anisotropy (low values of Q) and small domain sizes. Therefore, we are dealing with two limit situations: a) stray fields with high strengths and covering large and well defined areas and b) stray fields with low strengths and covering small and disordered areas.

The paper is organized as follows: In the results section, first we present the experimental results of sample A (Nb film/Co-Pd multilayer) and then the experimental results of sample B (Nb film/NdCo alloy). In the discussion section we analyze the experimental data for both samples. This discussion section is followed by a long summary and conclusions section where we present the most relevant outcomes. In the last section the fabrication and experimental methods are explained.

**Results**

**Sample A (Nb film/Co-Pd multilayer).** The array of Co/Pd triangles has been prepared in two different magnetic configurations. First, the sample has been measured in the "as grown" state, with remanence $M_R \approx 0$. This demagnetized state is reached in the usual way by minor hysteresis loops decreasing the strength of the applied magnetic field up to zero. A Magnetic Force Microscopy (MFM) image of the as grown state (Fig. 1(a)) shows that each triangle is broken up into a small number of black/white contrast regions (two or three) that correspond to large domains with opposite out-of-plane magnetization of sizes similar to the triangle dimensions very similar to multidomain states observed in CoPt dots [40]. Domain configuration is uncorrelated from triangle to triangle, indicative of a disordered magnetic state. This will be labelled as "Sample A - as grown". Second, the sample has been saturated with a strong out-of-plane $H_z$ magnetic field (30 kOe) and brought down to remanence (labelled as "sample A – $H_z$ remanence"). In this case, a more ordered configuration appears (see the remanent MFM image in Fig. 1(b) In this image magnetic contrast is weaker, corresponding to positive $M_z$ orientation in most triangles. A broad a white contrast region can be seen at the triangle centers, corresponding to a mostly out-of-plane magnetization state [28, 40]. The out-of-plane hysteresis loop of the patterned array of Co/Pd triangles presents a high value of the remanent magnetization $M_R = 0.85 M_S$, in agreement with the MFM characterization.

Material properties of Co/Pd multilayers are only weakly temperature dependent below room temperature (both saturation magnetization (41) and perpendicular magnetic anisotropy [42]). Thus, taking into account that the magnetic configuration of the Co/Pd triangles depends on the interplay between magnetostatic energy (given by shape and saturation magnetization), exchange and anisotropy energies that are only weakly temperature dependent, we can assume that room temperature magnetic configuration does not change significantly as temperature goes down to 4K. Therefore, given the weak temperature dependence of the magnetic properties of Co/Pd multilayers [41, 42], these magnetic configurations stay invariant as the temperature is lowered down to 4 K in order to study the behavior of the hybrid superconducting/magnetic system.

Magnetotransport measurements have been carried out in the hybrid sample Nb film on top of the array of triangles made with Co-Pd multilayers, at constant temperature close to the superconducting critical temperature, where the effect of the periodic pinning due to the array overcomes the random intrinsic pinning and the resistance minima appear [43 - 45] (see methods). Figure 2 shows the experimental results in the sample A - as grown. Remarkably, the resistance minima are absent. Figure 3 shows the magnetoresistance in the case sample A – $H_z$ remanence. Two experimental facts can be underlined: i) sharp and well-defined resistance minima appear (notice the log scale in Y-axis), evenly spaced at the expected value 36 Oe, obtained from B = n $\Phi_0$; $\Phi_0$ being the quantum fluxoid ($\Phi_0 = 2.07 \cdot 10^{-7}$ gauss cm$^2$) and n being the density of pinning centers; i.e. n = 1/(a x b) (a, b being the sides of the rectangle unit cell of the array). ii) The R(H) curves are symmetrically shifted from the zero magnetic field value, when the remanent magnetization ($M_z$) is changed from up to down.

Recently, del Valle et al. have reported two interesting findings on commensurability effects in hybrid samples fabricated with arrays in contact with superconducting films [46]. First, periodic roughness in the superconducting film is enough to generate resistivity minima. According to this result a matching effect should be observed in hybrid sample A-as grown (Nb film/[Co/Pd] multilayer triangles). In this sample the periodic roughness is obtained by the triangular array embedded in the Nb film. Therefore, the periodic array has to induce commensurability effect between the vortex lattice and the array and resistance minima should be observed. Figure 2 shows that this matching effect is absent. The second finding reported in [46] is that preserving the local order is crucial to achieve matching effects. If the local order is not retained the magnetoresistance minima vanish. These two findings are a hint to explain the lack of magnetoresistance minima in sample A-as grown. As seen in Fig. 1 (a) the demagnetized state of sample A shows nanomagnets with large areas of up and down stray fields but with different distribution from one triangle to another. The array of triangles does not show any local magnetic order at all. The effect of random distribution of the stray fields, which are different in each triangle, is strong enough to overcome the slowdown of the vortex motion (resistance minima) which should happen at certain magnetic field values (commensurability effect). Therefore, we can conclude that the geometrical matching effect is weaker than the effect of the random magnetic pinning potentials. In this case, the lack of local magnetic order suppress the effect induced in the vortex lattice dynamics by the periodic roughness on the hybrid sample

The resistivity minima appear when the sample is at the remanent state after saturation, see Fig. 3. The most remarkable effect is the asymmetry of the signal and the strong shift with the applied magnetic field. Morgan and Ketterson [47] reported an asymmetry in critical currents when the array of magnetic nanodots shows out-of-plane magnetization. Actually, Fig. 3 is a clear proof of the existence of magnetic-field-induced superconductivity [16-18]. In hybrid sample A the applied magnetic field enhances the critical temperature from $T_c$ (H=0) = 8.42 K to $T_c$ (H=240 Oe) = 8. 50 K, see [48], being 240 Oe the compensation field. Moreover, the superconducting vortex motion discriminates between upward (positive) or downward (negative) magnetization at remanence. The magnetoresistance shift to positive or negative values of the magnetic field is a good fingerprint of the remanent magnetization direction.

**Sample B (Nb film/NdCo alloy)** The magnetic configuration of sample B at different temperatures has been obtained from micromagnetic simulations performed with MuMax$^3$ code [49] as reported in [50]. At each temperature constant $M_S$ and $K_n$ values are used, obtained from the experimental characterization of single Nd-Co films [39, 50]. Material microstructure in Rare Earth-Transition Metal alloys is usually described in terms of random magnetic anisotropy models [51, 52]. Thus, to simulate microstructural disorder, the sample has been divided into a random set of exchange coupled "grains" of average size 12 nm and easy anisotropy axes randomly distributed around the average out-of-plane magnetization direction in a cone of width $\Delta\alpha$ (typically of the order of 15º). Figure 4(a) shows a micromagnetic calculation of the as-grown magnetict state of NdCo triangle (55 nm thick, 650 nm side) using room temperature (RT) material parameters $M_S$(RT) and $K_N$(RT). The triangle is divided in up/down (white/black) stripe domains that do not show any global preferred orientation due to the absence of an applied field. Only, near the borders, stripes tend to orient either parallel or perpendicular to the triangle edges, as expected for energy minimization [53]. Stripe width is quite regular, in the 50 nm – 80 nm range, almost an order of magnitude smaller than triangle size. These smaller domains are the result of magnetostatic energy minimization. They significantly reduce the stray field created by the magnetic triangles on the Nb film in the hybrid magnetic/superconducting sample both in intensity and in spatial range [31].

The low temperature magnetic configuration in "sample B-as grown" has been simulated starting with the configuration in Fig. 4(a) and gradually changing material parameters from the ones at room temperature to the ones at low temperature values, without any external applied field (i.e. gradually increasing $M_S$ from $M_S$(RT) to $M_S$(LT) and $K_N$ from $K_N$(RT) to $K_N$(LT)).

The result is shown in Fig. 4(b). The magnetic state at room temperature is frozen when the temperature is lowered; i. e. the global disorder at room temperature remains at low temperature. The only change obtained, decreasing temperature, is that domain walls between up and down domains become narrower due to the larger $K_N$. A fully different picture appears when the state "sample B – $H_z$ remanence" is simulated; see Fig. 4(c).

In this case, the sample is saturated, at low temperature, with a strong out of plane field and the magnetic configuration is simulated as the field is decreased down to remanence along a hysteresis loop. As can be seen, reversed magnetic domains are strongly disordered both in size and position within the triangle. This can be attributed to the lower mobility of the narrower low temperature domain walls that become more easily pinned at material defects.

Micromagnetic simulations in arrays with increasing degree of microstructural disorder give interesting insight into the different magnetic behavior of samples A and B at low temperatures (both of them with Q>1, i.e. stronger anisotropy than magnetostatic energy terms). Figure 5 shows the reduced remanent magnetization $M_R/M_S$ of triangles with Q = 1.4 and increasing $\Delta\alpha$, i.e. increasing disorder in easy anisotropy axes. For low $\Delta\alpha$, i.e. ordered material with well aligned easy axes, $M_R/M_S \approx 1$ and no reversed domains are observed at remanence (with a behavior equivalent to the high remanence Co/Pd alloys). Only for $\Delta\alpha$ above 10º, microstructural disorder becomes large enough to favor nucleation of reversed domains inside the triangle as the field is reduced from saturation to remanence (corresponding to the low remanence values of Nd-Co alloys).

In sample B (NdCo alloy) the behavior of the magnetoresistance clearly deviates from the behavior recorded in sample A (Co-Pd multilayer) in both states. Figure 6 shows the magnetoresistance in sample B-as grown state. Sharp resistance minima appear and they follow the well-known behavior reported in many hybrid systems [45] and references therein. That is, the periodicity of the minima is related to matching effects with the array. We have to remember that in sample A-as grown the minima are absent.

Figure 7 shows the magnetoresistance data in sample B – $H_z$ remanence. That is, the data are recorded after zero-field cooling up to 10 K and then applying an out of plane saturating magnetic field (30 kOe) and reducing the field to zero. Interestingly, in this remanent state of the NdCo alloy nanotriangles, the magnetoresistance curves in the hybrid sample follow an unexpected trend. The usual behavior is a strong enhancement of the pinning effects when the temperature is decreased [45] and references therein. In sample B – $H_z$ remanence the opposite happens; the matching effect vanishes decreasing the temperature.

**Discussion**

The main results can be summarized as follows: i) as grown states display a complete different behavior. In one case, strong anisotropy (sample A), matching effects are absent; but in the case of moderate magnetic anisotropy (sample B), matching effects are present and follow the usual behavior. ii) $H_z$ remanent state of the sample with strong anisotropy (sample A) gives rise to a R(H) curve strongly shifted, while in the case of moderate anisotropy (sample B) the minima are symmetric but they vanish decreasing the temperature.

We can explain this complex behavior taking into account the interplay between disorder and order of pinning magnetic potentials generated by stray magnetic fields, as well as, the competition between two characteristic length scales: the magnetic domain sizes and the superconducting vortex core dimension.

As Fig. 6 shows, sample B – as grown exhibits the usual behavior of the vortex dynamics on periodic arrays reported in the literature (see for instance in [45] and references therein). The superconducting vortex lattice matches with the array unit cell and well-defined magnetoresistance minima appear which are easier detected decreasing the temperature.

The most remarkable difference between sample A and sample B is that the latter shows a moderate out of plane magnetic anisotropy and consequently the distribution of stray fields up and down is weaker than in sample A (see Fig. 4). This smoother combination produces a less effective effect of the stray fields on the vortex dynamics than in sample A. So the magnetic contribution to matching effects is smaller than in sample A and the geometrical pinning potentials govern the vortex dynamics. Therefore, the vortex dynamics behaves as usual [45].

To shed light on these differences between these two systems an estimation, by micromagnetic simulations, of the stray field strength variations have been performed in two situations: magnetic stray fields on top of the triangles and out of the triangles. In sample A (Nb film/Co-Pd multilayer), with no external field and at a height z=50 nm (in the center of the Nb film), the calculation shows that the stray field is in the 600-1000 Oe range on top of the triangles, and around 250-350 Oe in the interstitial areas, among triangles. This means that each triangle generates a magnetic flux around 5 $\Phi_0$, enough to induce 5 vortex-antivortex pairs and in good agreement with the experimental shifted magnetoresistance and the results reported by del Valle et al. (48). In the case of sample B (Nb film/NdCo alloy) the stray field generated by a 55 nm thick NdCo film at the mid-plane (height z=50 nm) of the 100 nm Nb film is of the order of 300 Oe in average on top of each magnetic triangles domains and the stray field is 40 Oe in the interstitial region among triangles at 50 nm height (in the center of the Nb film). This value is too low to generate interstitial vortices (anti-vortices) in these regions, as can be seen by the zero-shift magnetoresistance curves, which means that there is not compensation field in this sample. In addition, the values of the stray fields are a clue to explain the magnetoresistance differences in the as-grown states. In sample A the disorder of strong stray fields overcomes the geometric order and the matching effect vanishes, conversely in sample B the smaller stray field strength does not preclude the matching effect and the periodic pinning effect induces the resistivity minima.

More interesting is the magnetoresistance behavior in the case of sample B – $H_z$ remanence (see Fig. 7). The most remarkable effect is that decreasing the temperature, the minima disappear. To explain this striking finding we have to deal with a new parameter: the size of the superconducting vortices. As was noticed in the introduction section, in principle, two mechanisms can be taken into account: i) the induced screening supercurrents needed to expel the stray magnetic fields and ii) the interplay between the coherence length (vortex core) and the magnetic domain or wall domain sizes. It turns out that the latter mechanism has been named by several authors [11-15] as the clue to explain the coexistence between superconductivity and magnetism. The vortex cores have to be commensurable with the magnetic domain sizes. Figure 7 shows a crossover from a regime in which resistance minima are observed to another regime in which minima are absent. This crossover happens decreasing the temperature. That is the vortex core shrinks. In our analysis, it is crucial to obtain an estimation of the vortex cores, i. e. the superconducting coherence lengths. The temperature dependence of the superconducting coherence length $\xi$ (T) can be obtained as usual from the measurement of the upper critical field temperature dependence, $H_{c2}$(T), see [54]. The temperature dependence of the coherence length $\xi$ (T) is plotted in Fig. 8. The shadow area represents the region where the coherence lengths are of the order or smaller than the domain sizes which can be estimated around 70 nm, see Fig. 4(c) and [37 - 39, 50]. Interestingly, the crossover temperature is similar to the temperature where the resistance minima disappear. This experimental result can be double checked measuring the temperature dependence of the critical currents. Minima in resistances correspond to maxima in critical currents. Inset in Fig. 8 shows $I_c$ (H) at constant temperature in the relevant temperature interval. We can observe the same temperature crossover that is found in the magnetoresistance (Fig. 8 inset). Hence, we can conclude that the vortices with core dimensions larger than the domain sizes average over the magnetic landscapes and they only follow the geometrical matching effects; but when the vortex cores shrink they begin to interact with the random magnetic domains and the magnetoresistance minima vanish.

It is interesting to note that superconducting vortices are sensitive to the subtle differences in disorder between as grown and $H_z$-remanence states: minima only vanish due to the disorder in the magnetic state for the latter state, in which magnetic domains are disordered both in size and position. In the as grown state, that retains certain degree of uniformity in domain size, minima are still present.

**Conclusions**

Our work deals with the interplay between commensurate pinning (created by the periodic triangle array) and order-disorder magnetic pinning created by the domain structure of two different materials. Two different length scales are involved in this order-disorder competition that are analyzed using either Co-Pd multilayer or NdCo alloy.

1) <u>Disorder on a length scale comparable with nanomagnet size and array unit cell dimensions.</u> This is achieved with a Co/Pd sample in the as-grown state: there are only one or two domains per triangle with sizes of several hundreds of nm, i.e. comparable to array unit cell (800 nm): Previous works [46] had shown that physical displacement of the pinning sites on a length scale comparable with the array length scale quench periodic pinning and minima in the resistivity disappear. The results for sample A (Fig. 1a and Fig. 2) show an alternative method to create disorder on this same length scale (using the domains within each Co/Pd triangle) but with a clear advantage over the method used in [46]: Magnetic disorder is erasable with a magnetic field (as shown in Fig. 3) in contrast with topographic disorder that is fixed once the sample is fabricated. Magnetic domain size also governs the spatial extension of the stray field in the superconductor. Single domain state comparable to element size (Fig. 3) creates an effective magnetic field on the Nb films that displaces the magnetoresistance curve either to positive or negative field values.

2)      Disorder on a length scale smaller than nanomagnets size and array unit cell dimensions (but comparable to vortex size at low temperatures).

The different behavior between sample A (Pd-Co multilayer) and sample B (NdCo alloy) clearly show that domain size is an important factor in magnetic pinning effect. Three important novel experimental facts are observed:

a.      Small magnetic domains do not create asymmetries in the magnetoresistance curve (neither in the as-grown or remanent states). This is a direct consequence of small mixed up/down domains that minimize magnetostatic energy so that the stray field barely extends beyond the magnetic triangles.

b.      Periodic pinning depends on the averaging of the effective magnetic pinning potential over vortex size (with a crossover when vortex size is of the order of magnetic domains). If the magnetic domain size is smaller than the coherence length (vortex core) the local order parameter in the superconductor cannot follow such rapid spatial modulations and the vortex "sees" an uniform order parameter, i.e. there is no pinning (as pinning depends on the gradients of the order parameter in the superconductor [55]). This is in agreement with earlier global measurements [17] and locally images by STM in [56].

c.      Subtle differences in magnetic disorder result in very different superconducting pinning behavior: 1) as-grown NdCo triangles display disordered stripes frozen from the high temperature configuration (Q = 0.3 at RT) in which orientation order is lost but there is a certain order in stripe domain position (stripe width is fixed and stripes tend to meet triangle borders either parallel or perpendicular) and 2) $H_z$-remanent NdCo triangles with magnetic disorder both in size and domain position/orientation. This second kind of disordered magnetic state is able to compete effectively with periodic pinning (no minima are observed at low temperatures) as long as vortex size is small enough to interact with the smaller domains in each triangle.

Finally, we want to stress that superconducting vortex dynamics can discriminate among magnetic states in arrays of nanomagnets: different remanent states of the magnetic arrangement of nanomagnets and subtle magnetic differences produce qualitatively different responses in the superconducting vortex dynamics since superconducting vortex lattice motion probes both the local magnetic state of nanomagnets and the ordered or disordered magnetic state in array of nanomagnets.

**Methods**

The periodic arrays of magnetic triangles have been fabricated by e-beam lithography and lift off on Si (100) substrates, as reported before, see for instance [48]. The equilateral triangles (650 nm sides) are arranged following a rectangular unit cell of 800 nm × 700 nm. The two different magnetic systems [Co/Pd multilayer (sample A) and NdCo alloy (sample B)] have been deposited by dc magnetron sputtering on Si (100) substrates, see details in [29] and [37 - 39] respectively. On top of these arrays, Nb film was grown by magnetron sputtering with 100 nm thickness, with a base pressure of $5\times10^{-8}$ Torr. Finally, for electric transport measurements, an 8-terminal cross-shaped bridge was defined using optical lithography and Ar/SF$_6$ (1:2) Reactive Ion Etching. The area of the bridge is 40 μm x 40 μm. In sample A the magnetic layer is a {Co [0.4 nm]/Pd [0.6 nm]} with total thickness 40 nm, deposited by sputtering from single Co and Pd targets. This Co/Pd multilayer is a magnetic system with perpendicular anisotropy and with saturation magnetization $M_s$ (RT) = 5 x$10^5$ A/m (58). Co and Pd layer thicknesses and sputtering pressure during growth were selected to optimize out-of-plane remanent magnetization. In a Co/Pd multilayer film of similar composition $M_R$ is 0.95$M_S$ (obtained from an out-of-plane hysteresis loop measured with a SQUID magnetometer at 5 K), characteristic of very strong perpendicular magnetic anisotropy (PMA) as was reported in the literature [29, 32-36]. In sample B, the magnetic layer is a Nd$_{16}$Co$_{84}$ film with 55 nm thickness, deposited by sputtering from Nd-Co target of a similar composition [50]. Nd$_{16}$Co$_{84}$ is a soft ferromagnetic alloy with $M_S$(RT)=7×$10^5$ A/m and PMA $K_N$(RT)=$10^5$ J/m$^3$ at room temperature (RT), i.e. Q(RT) = 0.3 corresponding to a weak PMA material. Equilibrium domain configuration consists of parallel stripes with alternating up-down magnetization and width in the 50-60 nm range [37-39] depending on sample thickness [39, 57]. In the demagnetized state, stripe domains become branched and curved in a labyrinth configuration [37] but they retain its characteristic width given by the competition between magnetostatic and anisotropy energies. At low temperatures (LT) (10 K), the anisotropy in the material increases up to $K_N$(LT)=1.8 × $10^6$ J/m$^3$ at 10 K [37, 38], with $M_S$(LT)=1.4×$10^6$ A/m. Thus, at low temperature Q increases up to Q = 1.4 which corresponds to a material with moderate anisotropy. In NdCo films of similar composition, out-of-plane remanence at low temperature presents relatively low values, of the order of 0.1 $M_S$ [37, 38], characteristic of strong disorder that favors reversed domain nucleation at positive fields and a disordered multidomain state at remanence.

Superconducting vortex motion is the proposed "experimental technique" to detect the magnetic states of these systems. Magneto-transport R(H) experiments, R being the resistance, were done with a magnetic field H applied perpendicular to the substrate in a liquid helium system with superconducting magnet [48]. The applied magnetic field does not modify the remanent magnetic state of the samples. The vortex lattice is set in motion by a driven current applied perpendicular to the base of the triangles. The dc magnetoresistance exhibits commensurability effects [43 - 45] in which dissipation minima develop as a consequence of the geometrical matching between the vortex lattice and the underlying periodic structure. At these matching fields, the vortex-lattice motion slows down, and R(H) sharp minima appear at well-defined and equally spaced values of the applied field H. The first matching field is the one for which the density of vortex lattice equals the density of pinning centers. We have chosen triangular shape for the nanomagnet instead the usual circular dots, since equilateral triangles are a good reference for selecting the driving current direction and the vortex motion, as well as the magnetic stripe domain directions, since stripes prefer to meet triangle borders either parallel or perpendicular [40].

References

(1) Ginzburg, V. L. Ferromagnetic Superconductors. *Zh. Eksp. Teor. Fiz*. **31**, 202 (1956) [*Sov. Phys. JETP* **4**, 153 (1957)].
(2) Matthias, B. T., Suhl, H. & Corenzwit, E. Spin exchange in superconductors. *Phys. Rev. Lett.* **1**, 92 (1958).
(3) Anderson P. W. & Suhl, H. Spin alignment in the superconducting state. *Phys. Rev*. **116**, 898 (1959).
(4) Maple, M. B. Superconductivity: A probe of the magnetic state of local moments in metals *Appl. Phys*. **9**, 179 (1976).
(5) Fertig, W. A., Johnston, D. C., De Long, L. E., Mccallum, R. W., Maple M. B. & Matthias, B. T. Destruction of superconductivity at the onset of long-range magnetic order in the compound $ErRh_4B_4$. *Phys. Rev. Lett.* **38**, 387 (1977).
(6) Lynn, J. W. *et al.* Temperature-dependent sinusoidal magnetic order in the superconductor $HoMo_6Se_8$. *Phys. Rev. Lett.,* **52**, 133 (1984).
(7) Bernhard, C. *et al.* Coexistence of ferromagnetism and superconductivity in the hybrid ruthenate-cuprate compound $RuSr_2GdCu_2O_8$ studied by muon spin rotation and dc magnetization *Phys. Rev. B* **59**, 14099 (1999)
(8) Saxena, S. S. *et al.* Superconductivity on the border of itinerant-electron ferromagnetism in $UGe_2$. *Nature (London)* **406**, 587 (2000).
(9) Stewart, G. R. Superconductivity in iron compound. *Rev. Mod. Phys.* **83**, 1589 (2011).
(10) Bert, J. A. *et al.* Direct imaging of the coexistence of ferromagnetism and superconductivity at the $LaAlO_3/SrTiO_3$ interface *Nat. Phys*. **7**, 767 (2011)
(11) Matthias B. T. & Suhl, H. Possible explanation of the "coexistence" of ferromagnetism and superconductivity. *Phys. Rev. Lett.* **4**, 51 (1960).
(12) Tachiki, M., Kotani, A. H., Matsumoto, H. & Umezawa, H. Superconducting Bloch-wall in ferromagnetic superconductors. *Solid St. Commun*. **32**, 599 (1979).
(13) Kulic, M. L. Importance of RKKY exchange interaction in the formation of a superconducting Bloch wall. *Phys. Lett. A* **83**, 46 (1981).
(14) Buzdin, A. I., Bulaevskii, L. N. & Panyukov, S. V. Existence of superconducting domain walls in ferromagnets. *Zh. Eksp. Teor. Phys*. **87**, 299 (1984) [*Sov. Phys. JETP* **60**, 174 (1984)].
(15) Buzdin, A. I. & Mel'nikov, A. S. Domain wall superconductivity in ferromagnetic superconductors. *Phys. Rev. B* **67**, 020503(R) (2003).
(16) Lange, M. J., Van Bael, M. J., Bruynseraede, Y. & Moshchalkov V. V. Nanoengineered magnetic-field-induced superconductivity. *Phys. Rev. Lett.* **90**, 197006 (2003).
(17) Yang, Z. R., Lange, M., Volodin, A., Szymczak, R. & Moshchalkov V. V. Domain-wall superconductivity in superconductor-ferromagnet hybrids. *Nat. Mat.* **11**, 793 (2004).
(18) Steiner R. & Ziemann P. Magnetic switching of the superconducting transition temperature in layered ferromagnetic/superconducting hybrids: Spin switch versus stray field effect. *Phys. Rev. B* **74**, 094504 (2006).
(19) Baert, M., Metlushko, V., Jonckheere, R., Moshchalkov, V. V. & Bruynseraede, Y. Composite flux-line lattices stabilized in superconducting films by a regular array of artificial defects. *Phys. Rev. Lett.* **74**, 3269 (1995).

(20) Martin, J. I., Velez, M., Hoffmann, A., Schuller, I. K. & Vicent J. L. Artificially induced reconfiguration of the vortex lattice by arrays of magnetic dots. *Phys. Rev. Lett.* **83**, 1022 (1999).

(21) Villegas, J. E. *et al.* A superconducting reversible rectifier that controls the magnetic flux quanta. *Science* **302**, 1188 (2003)

(22) V. Vlasko-Vlasov, *et al.* Guiding superconducting vortices with magnetic domain walls. *Phys. Rev. B* **77**, 134518 (2008)

(23) Milosevic, M. V. & Peeters F. M. Vortex-antivortex lattices in superconducting films with magnetic pinning arrays *Phys. Rev. Lett.* **93**, 267006 (2004)

(24) Gomez, A., Gonzalez, E. M., Gilbert, D. A., Milosevic, M. V., Liu, K. & Vicent, J. L. Probing the dynamic response of antivortex, interstitial and trapped vortex lattices on magnetic periodic pinning potentials. *Supercon. Sci. Technol*. **26**, 085018 (2013)

(25) Bobba, F. *et al.* Vortex-antivortex coexistence in Nb-based superconductor/ferromagnet heterostructures. *Phys. Rev. B* **89**, 214502 (2014)

(26) Belkin, A., Novosad, V., Iavarone, M., Pearson, J. & Karapetrov, G. Superconductor/ferromagnet bilayers: Influence of magnetic domain structure on vortex dynamics. *Phys. Rev. B* **77**, 180506(R) (2008).

(27) Moore, S. A. *et al.* Doppler-scanning tunneling microscopy current imaging in superconductor-ferromagnet hybrids. *Appl. Phys. Lett*. **108**, 042601 (2016).

(28) Martín, J. I., Nogués, J., Liu, K., Vicent, J. L. & Schuller Ivan K. Ordered Magnetic Nanostructures: Fabrication and Properties. *J. Magn. Magn. Mater*. **256**, 449 (2003).

(29) Gomez, A., Gilbert, D. A., Gonzalez, E. M., Liu, K. & Vicent, J. L. Control of dissipation in superconducting films by magnetic stray fields. *Appl. Phys. Lett*. **102**, 052601 (2013)

(30) Jagla, E. A. Hysteresis loops of magnetic thin films with perpendicular anisotropy. *Phys. Rev. B* **72**, 094406 (2005)

(31) Huber, A. & Schäfer, R. *Magnetic Domains* (Springer-Verlag, Berlin-Heidelberg 1998).

(32) Hashimoto, S., Ochiai, Y. & Aso, K. Perpendicular Magnetic Anisotropy in Sputtered CoPd Alloy Films. *Jpn. J. Appl. Phys*. **28**, 1596 (1989).

(33) de Haan, P., Meng, Q., Katayama, T. & Lodder, J. C. Magnetic and magneto-optical properties of sputtered Co/Pd multilayers. *J. Magn. Magn. Mater*. **113**, 29 (1992)

(34) Ngo, D.-T., *et al*. Interfacial tuning of perpendicular magnetic anisotropy and spin magnetic moment in CoFe/Pd multilayers. *J. Magn. Magn. Mater*. **350,** 42 (2014).

(35) Kirby, B. J., *et al*. Direct observation of magnetic gradient in Co/Pd pressure-graded media. *J. Appl. Phys*. **105**, 07C929 (2009).

(36) Stärk, M. *et al*. Controlling the magnetic structure of Co/Pd thin films by direct laser interference patterning. *Nanotechnology* **26**, 205302 (2015)

(37) Ruiz-Valdepeñas, L. *et al.* Double percolation effects and fractal behavior in magnetic/superconducting hybrids. *New J. Phys.* **15**, 103025 (2013)

(38) Ruiz-Valdepeñas, L. et al. Imprinted labyrinths and percolation in Nd-Co/Nb magnetic/superconducting hybrids. *J. Appl. Phys.* **115**, 213901 (2014)

(39) Hierro-Rodriguez, A. *et al*. Fabrication and magnetic properties of nanostructured amorphous Nd–Co films with lateral modulation of magnetic stripe period. *J. Phys. D: Appl. Phys.* **46,** 345001 (2013).

(40) Abes, M. et al. Magnetic switching field distribution of patterned CoPt dots. *J. Appl. Phys.* **105** 113916 (2009)

(41) Draaisma, H. J. G., den Broeder F. J. A. & de Jonge, W. J. M. Perpendicular anisotropy in Pd/Co multilayers. *J. App. Phys*. **63**, 3479 (1988).

(42) Hong, J. I., Sankar, S., Berkowitz, A. E. & Egelhoff Jr., W. F. On the perpendicular anisotropy of Co/Pd multilayers. *J. Magn. Magn. Mat.* **285**, 359 (2005).

(43) Martín, J. I., Vélez, M., J. Nogués, J. & Schuller, I. K. Flux Pinning in a Superconductor by an Array of Submicrometer Magnetic Dots. *Phys. Rev. Lett.* **78**, 1929 (1997).

(44) Reichhardt, C., Olson, C. J. & Nori, F. Commensurate and incommensurate vortex states in superconductors with periodic pinning arrays. *Phys. Rev. B* **57**, 7937 (1998).

(45) Vélez, M. et al. Superconducting vortex pinning with artificial magnetic nanostructures. *J. Magn. Magn. Mat.* **320**, 2547 (2008) and references therein.

(46) del Valle, J., Gomez, A., Luis-Hita, J., Rollano, V., Gonzalez, E. M. & Vicent, J. L. Different approaches to generate matching effects using arrays in contact with superconducting films. *Supercon. Sci. Technol.* **30**, 025014 (2017).

(47) Morgan D. J. & Ketterson, J. B. Asymmetric Flux Pinning in a Regular Array of Magnetic Dipoles. *Phys. Rev. Lett*. **80**, 3614 (1998).

(48) del Valle, J., Gomez, A., Gonzalez, E. M., Osorio, M. R., Granados, D. & Vicent, J. L. Superconducting/magnetic three-state nanodevice for memory and reading applications. *Sci. Rep*. **5**, 15210 (2015).

(49) Vansteenkiste, A., Leliaert, J., Dvornik, M., Helsen, M., Garcia-Sanchez, F. & Van Waeyenberge, B. The design and verification of MuMax3. *AIP Advances* **4**, 107133 (2014).

(50) Hierro-Rodriguez, A., et al. Observation of asymmetric distributions of magnetic singularities across magnetic multilayers. *Phys. Rev. B* **95**, 014430 (2017).

(51) Alameda, J. M. et al. Co anisotropy in amorphous $Y_{1-x}Co_x$ films. *J. Magn. Magn. Mater*. **67** 115 (1987);

(52) Alben, R., Becker, J. J. & Chi, M. C. Random anisotropy in amorphous ferromagnets. *J. Appl. Phys.* **49** 1653 (1978)

(53) Clarke, D., Tretiakov, O. A., & Tchernyshyov, O. Stripes in thin ferromagnetic films with out-of-plane anisotropy. *Phys. Rev. B* **75**, 174433 (2007).

(54) Tinkham, M. *Introduction to Superconductivity* (McGraw-Hill Inc., New York, 1996).

(55) Campbell A. M. & Evetts, J. E. Flux vortices and transport currents in type II superconductors. *Adv. Phys.* **21**, 199 (1972).

(56) Iavarone, M. *et al.* Visualizing domain wall and reverse domain superconductivity. *Nat. Comm*. **5**, 4766 (2014).

(57) Carcia, P. F., Meinhaldt, A. D. & Suna, A. Perpendicular magnetic anisotropy in Pd/Co thin film layered structures. *Appl. Phys. Lett.* **47**, 178 (1985).

(58) Blanco-Roldán, C., et al. Nanoscale imaging of buried topological defects with quantitative X-ray magnetic microscopy. *Nat. Comm.* **6**, 8196 (2015).

**Acknowledgments**

We thank Spanish MINECO grants FIS2013-45469, FIS2016-76058 (AEI/FEDER, UE),  Spanish CM grant S2013/MIT-2850. IMDEA Nanociencia acknowledges support from the 'Severo Ochoa' Programme for Centres of Excellence in R&D (MINECO, Grant SEV-2016-0686). DG acknowledges RYC-2012-09864, S2013/MIT-3007 and ESP2015-65597-C4-3-R for financial support.

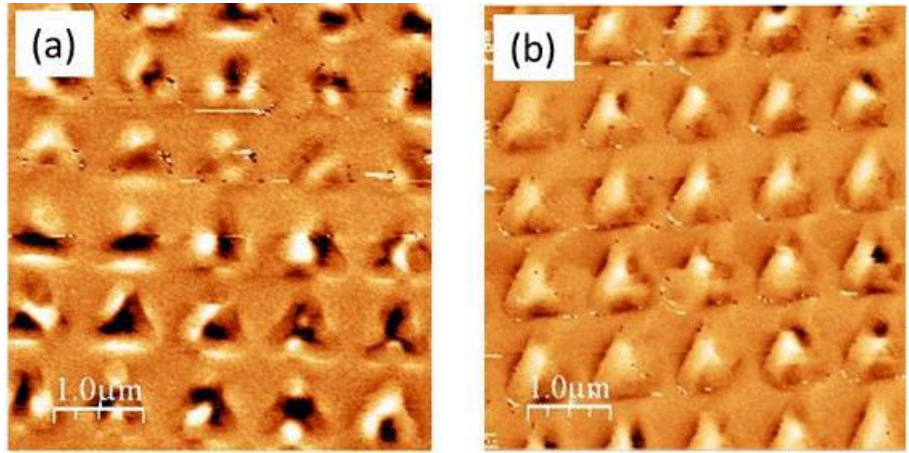

FIG. 1. MFM image of Co/Pd triangles at 300 K: (a) as grown; (b) at remanence after out-of-plane saturation.

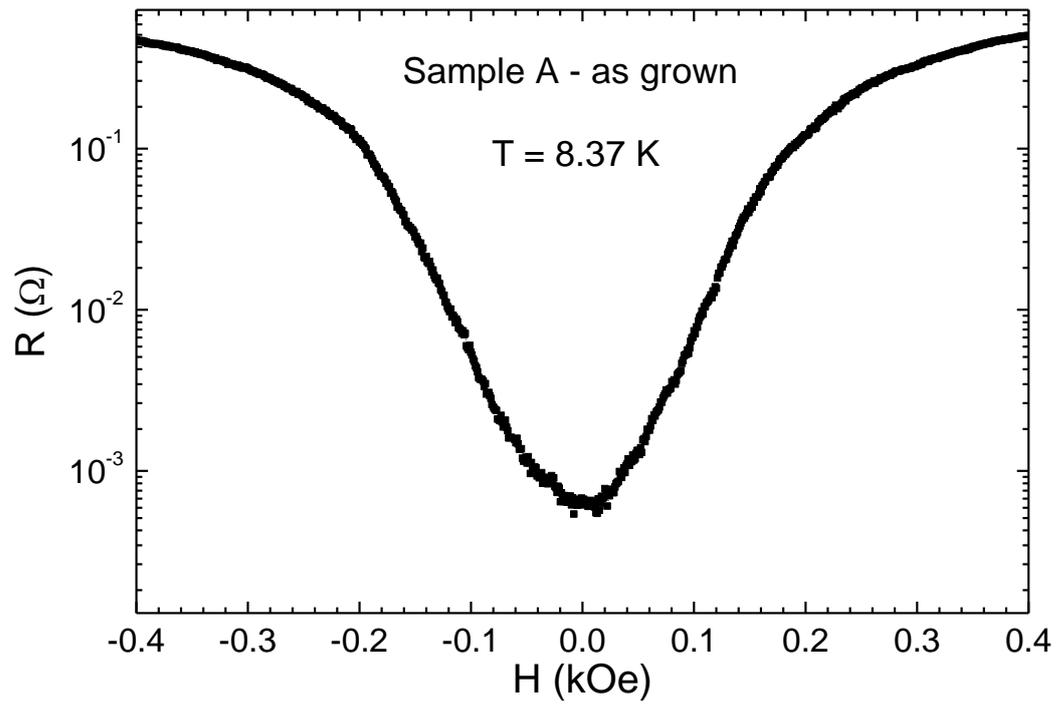

FIG. 2. Magnetoresistance of sample A-as grown (Nb film/[Co/Pd] multilayer triangles) $T_c(H=0)$ = 8.42 K.

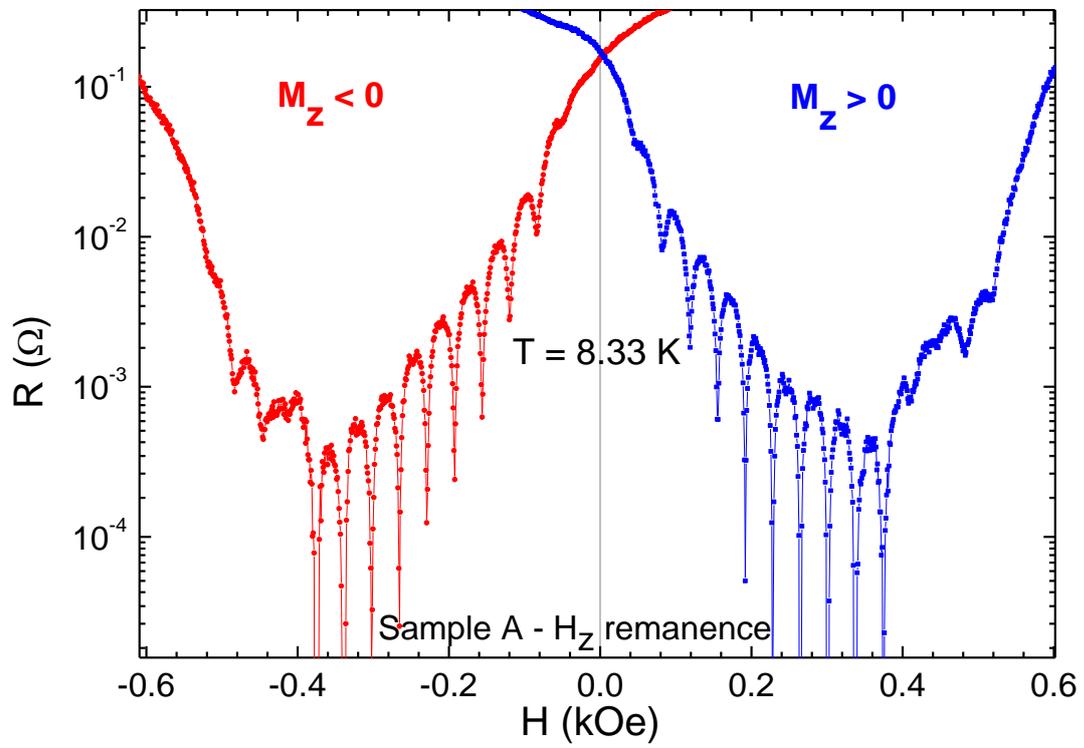

FIG. 3. Magnetoresistance of sample A – $H_z$ remanence (Nb film/[Co/Pd] multilayer triangles) $T_c(H=0) = 8.42$ K.

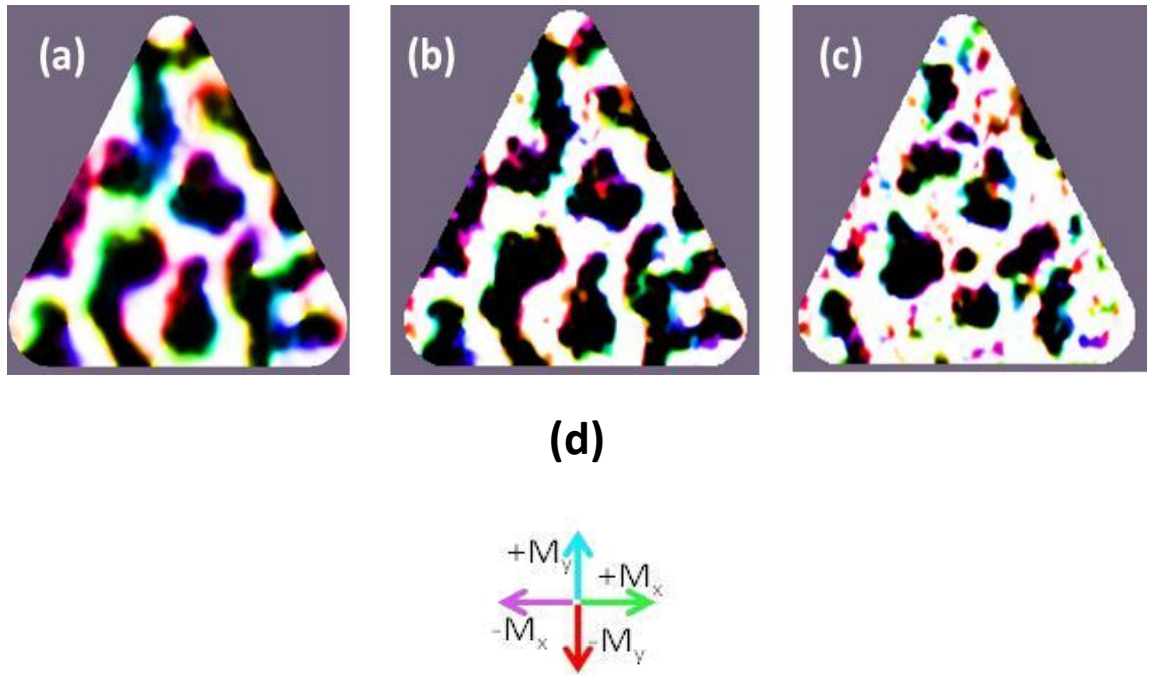

FIG. 4. Micromagnetic simulations of NdCo triangles: (a) at room temperature with Q(RT)=0.3 starting from a disorder state at zero field; (b) at low temperature with Q(LT)=1.4 starting from state in (a) at zero field; (c) at low temperature with Q(LT)=1.4 starting from an out-of-plane saturated state in high field ($\mu_0 H_z$= 3 T) and reducing H down to zero along a hysteresis loop. Images show the magnetization configuration at the central plane of the triangle, black/white contrast corresponds to $-M_S/+M_S$ out of plane magnetization, respectively. The size of the triangle is exactly the same that the fabricated NdCo triangle. Note that domain boundaries are Bloch walls inside the sample, see (d) arrow sketch, with small Neel caps near the surface [31]. Walls become sharper as temperature is reduced.

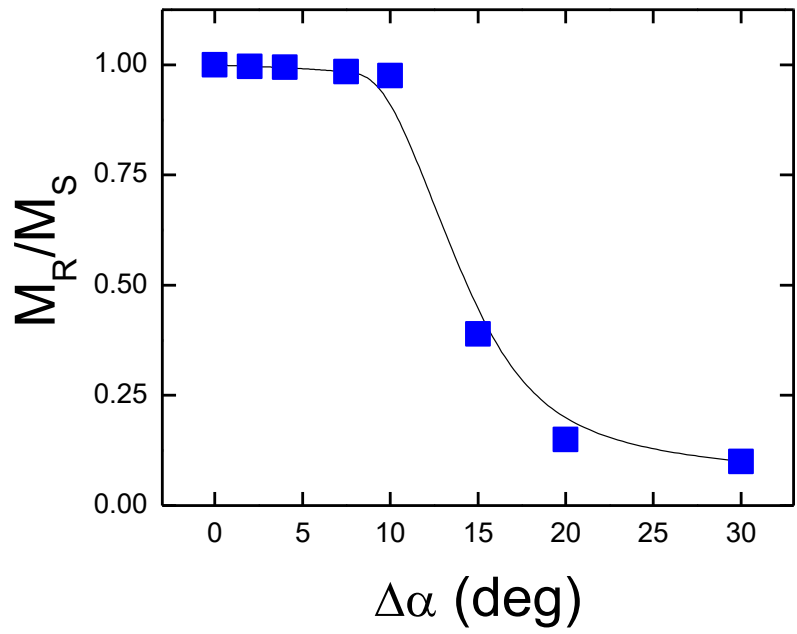

FIG. 5. Remanent magnetization of Nd-Co triangles (LT Q=1.4) as a function of magnetic disorder (Δα is the angular width of the random distribution of easy anisotropy axes). Reversed domains are only observed for Δα above 10º. The line is a guide to the eye.

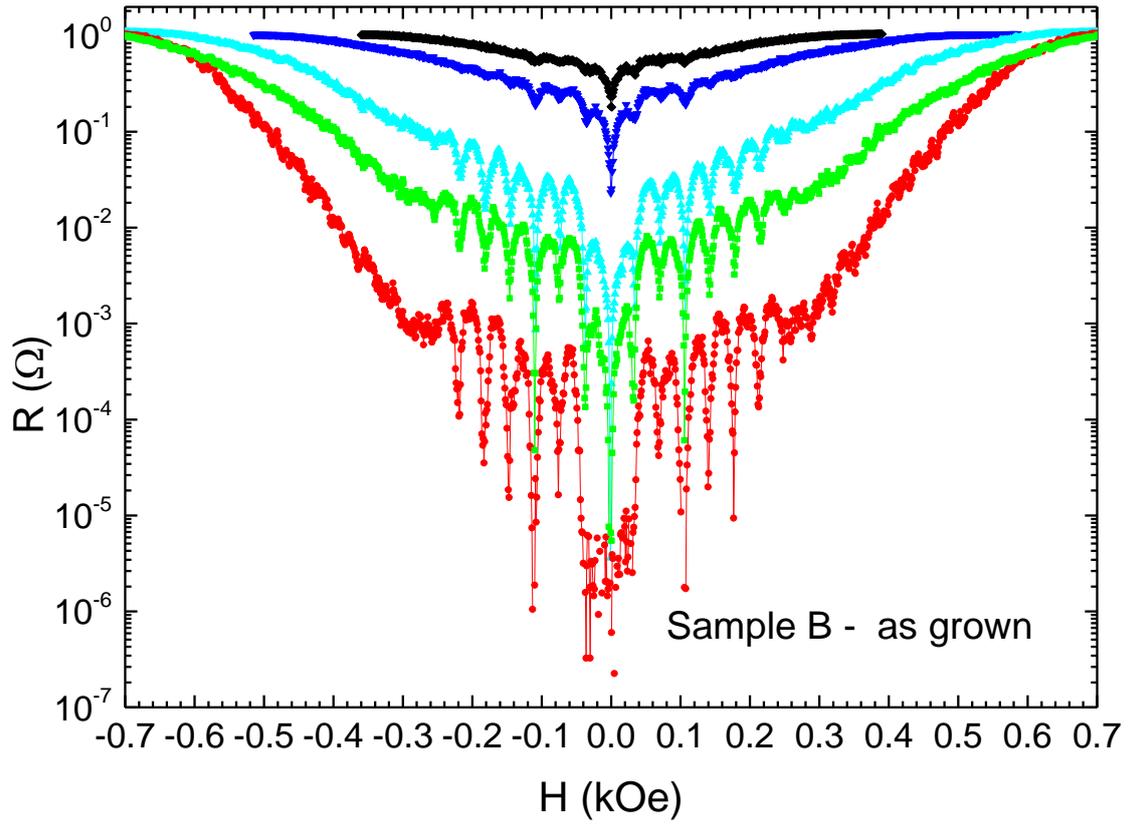

FIG. 6. Magnetoresistance of sample B-as grown (Nb film/NdCo alloy triangles). Temperatures from up to down curves: 8.57 K, 8.53 K, 8.50 K, 8.44 K and 8.35 K. $T_c$ (H=0) = 8.61 K.

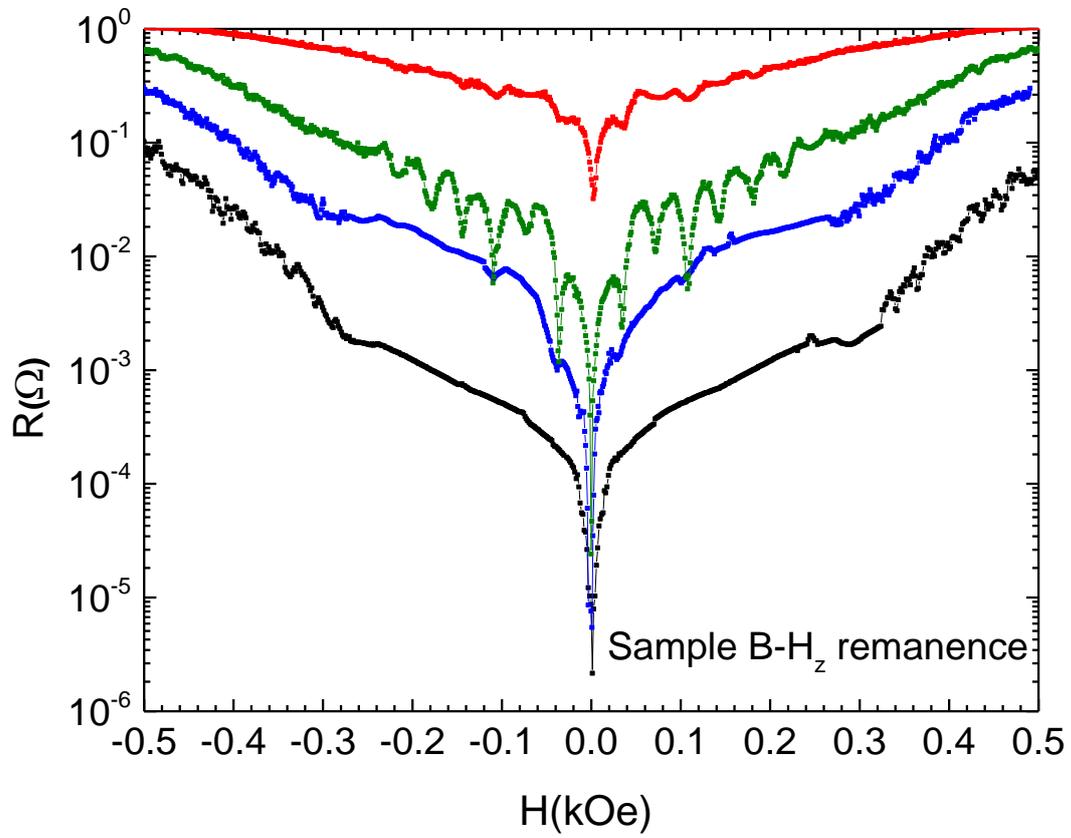

FIG. 7. Magnetoresistance measurements for sample B – $H_z$ remanence state (Nb film/NdCo alloy triangles). Temperatures from up to down curves: 8.53 K, 8.48 K, 8.43 K and 8.35 K. $T_c$ (H=0) = 8.61 K.

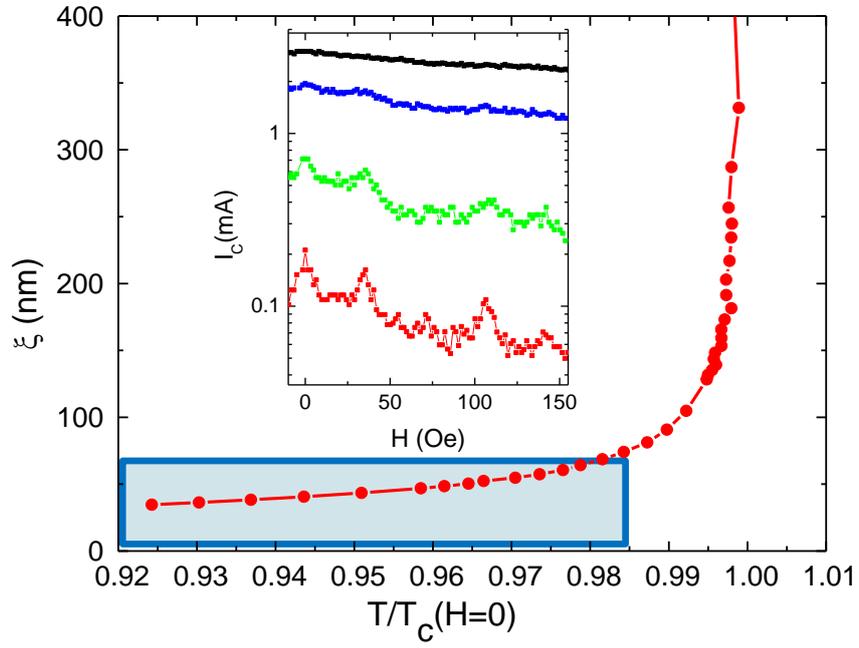

FIG. 8. Sample B (Nb film/NdCo alloy). Y-axis coherence length ($\xi$). X-axis reduced temperature $T_c(H=0)$ = 8.61 K. Shading area shows lengths smaller than the sizes of the magnetic domains. Inset: Sample B (Nb film/NdCo alloy). Y-axis critical currents ($I_c$). X-axis applied magnetic fields. Temperatures from down to up curves: 8.56 K, 8.52 K, 8.43 K and 8.35 K. $T_c(H=0)$ = 8.61 K. Voltage criterion to obtain $I_c$ is V = 3 x 10$^{-7}$ V.